\documentstyle[twoside,fleqn,espcrc2,epsf]{article}
\title{    Where is the string limit in QCD?\thanks{Talk
           presented by C.~Morningstar and poster
           presented by K.J.~Juge.}}

\author{K.J.~Juge, J.~Kuti, and C.J.~Morningstar\address{Dept.~of Physics,        University of California at San Diego,
        La Jolla, California 92093-0319}}

\begin{document}

\begin{abstract}

The energies of glue in the presence of a static quark-antiquark pair
are calculated for separations $r$ ranging from 0.1 fm to 4 fm and for
various quark-antiquark orientations on the lattice.  Our simulations
use an improved gauge-field action on anisotropic space-time lattices.
Discretization errors and finite volume effects are studied.  We find
that the spectrum does not exhibit the expected onset of the universal
$\pi/r$ Goldstone excitations of the effective QCD string, even for $r$
as large as 4 fm.  Our results cast serious doubts on the validity of
treating glue in terms of a fluctuating string for $r$ below 2 fm.
Retardation effects in the $\Upsilon$ system are also studied by
comparing level splittings from the Born-Oppenheimer approximation with
those directly obtained in simulations.

\end{abstract}
\maketitle

\section{Introduction}

Accurate knowledge of the properties of the stationary states of
glue in the presence of the simplest of color sources, that of a static
quark and antiquark separated by some distance $r$, is an
important stepping stone on the way to understanding confinement.
It is generally believed that at large $r$, the linearly-growing
ground-state energy of the glue is the manifestation of the confining
flux whose fluctuations can be described in terms of an effective
string theory.  The lowest-lying excitations are then
the Goldstone modes associated with spontaneously-broken transverse
translational symmetry.  Expectations are less clear for small $r$.

Even the simplest property, the energy spectrum, of the stationary
states of glue interacting with a static quark-antiquark pair is not
accurately known.  The main goal of this work is to remedy this.  Here,
we present, for the first time, a comprehensive determination of the
low-lying spectrum of gluonic excitations in the presence of a static
quark-antiquark pair.  In this initial study, the effects of light
quark-antiquark pair creation are ignored.  A few of the energy levels for
$r$ less than 1 fm have been studied before\cite{previous}.  Our results
for these quantities have significantly improved precision, and we
have extended the range in $r$ to 4 fm.  Most of the energy levels
presented here have never been studied before.  Some of our results
were previously reported\cite{earlier}. 

The determination of the energies of glue in the presence of a static
quark-antiquark pair is also the first step in the Born-Oppenheimer
treatment of conventional and hybrid heavy-quark mesons\cite{hasenfratz}.
The validity of the Born-Oppenheimer expansion depends, in part, on the
smallness of retardation effects.  In order to quantify such effects
for the first time,  mass splittings
for various conventional and hybrid heavy-quark mesons obtained from the
leading Born-Oppenheimer approximation are compared to those determined 
in simulations in which the heavy quark propagates according to a
spin-independent nonrelativistic action.  

\section{Computation of the glue energies}

We adopt the standard notation from the physics of diatomic molecules
and use $\Lambda$ to denote the magnitude of the eigenvalue of the projection
$\vec{J_g}\!\cdot\hat{\bf r}$ of the total angular momentum $\vec{J_g}$
of the gluons onto the molecular axis $\hat{\bf r}$. The capital Greek
letters $\Sigma, \Pi, \Delta, \Phi, \dots$ are used to indicate states
with $\Lambda=0,1,2,3,\dots$, respectively.  The combined operations of
charge conjugation and spatial inversion about the midpoint between the
quark and the antiquark is also a symmetry and its eigenvalue is denoted by
$\eta_{CP}$.  States with $\eta_{CP}=1 (-1)$ are denoted
by the subscripts $g$ ($u$).  There is an additional label for the
$\Sigma$ states; $\Sigma$ states which
are even (odd) under a reflection in a plane containing the molecular
axis are denoted by a superscript $+$ $(-)$.  Hence, the low-lying
levels are labelled $\Sigma_g^+$, $\Sigma_g^-$, $\Sigma_u^+$, $\Sigma_u^-$,
$\Pi_g$, $\Pi_u$, $\Delta_g$, $\Delta_u$, and so on.  For convenience,
we use $\Gamma$ to denote these labels in general.

\begin{table}[t]
\setlength{\tabcolsep}{1mm}
\caption[tabone]{Simulation parameters, including the coupling $\beta$,
  input aspect ratio $\xi$, lattice size, and the spatial link 
  smearing parameters $\zeta$ and $n_{\zeta}$.  The approximate lattice
  spacings $a_s$, calculated assuming $r_0^{-1}=410$ MeV, are also given.
  \label{table:simparams}}
\begin{center}
\begin{tabular}{@{\hspace{0mm}}cccccc@{\hspace{0mm}}} \hline
 Run & $a_s$ (fm) & $\beta$ & $\xi$  & Lattice & $(\zeta, n_{\zeta})$
   \\ \hline\hline
A & 0.29 & 2.1 & 8 & $(10^2\!\times\!20)\times\!80$ & (0.07, 12)\\
  &      &     &   &                                & (0.10, 12)\\
  &      &     &   &                                & (0.15, 12)\\ \hline
B & 0.29 & 2.1 & 8 & $(10^2\!\times\!20)\times\!80$ & (0.10, 12)\\ \hline
C & 0.29 & 2.1 & 8 & $14^3\!\times\!56$             & (0.15,  6)\\
  &      &     &   &                                & (0.10, 12)\\ \hline
D & 0.27 & 2.2 & 5 & $12^3\!\times\!48$             & (0.10,  4)\\
  &      &     &   &                                & (0.20,  4)\\
  &      &     &   &                                & (0.30,  4)\\ \hline
E & 0.22 & 2.4 & 5 & $14^3\!\times\!56$             & (0.10,  8)\\
  &      &     &   &                                & (0.15,  8)\\
  &      &     &   &                                & (0.25,  8)\\
  &      &     &   &                                & (0.30,  8)\\ \hline
F & 0.19 & 2.5 & 5 & $(10^2\!\times\!25)\times\!60$ & (0.10, 12)\\
  &      &     &   &                                & (0.18, 12)\\
  &      &     &   &                                & (0.26, 12)\\ \hline
G & 0.19 & 2.6 & 3 & $10^3\!\times\!30$             & (0.15,  8)\\
  &      &     &   &                                & (0.30,  8)\\ \hline
H & 0.12 & 3.0 & 3 & $15^3\!\times\!45$             & (0.15, 24)\\
  &      &     &   &                                & (0.22, 24)\\ \hline
\end{tabular}
\end{center}
\end{table}

The glue energies $E_\Gamma(\vec{r})$ were extracted from Monte Carlo
estimates of generalized Wilson loops.  Recall that the well-known
static potential $E_{\Sigma_g^+}(r)$ can be obtained from the large-$t$
behaviour $\exp[-tE_{\Sigma_g^+}(r)]$ of the Wilson loop for a rectangle
of spatial length $r$ and temporal extent $t$.  In order to determine the
lowest energy in the $\Gamma$ sector, each of the two spatial segments of the
$r\times t$ rectangular Wilson loop must be replaced by a {\em sum} of
spatial paths, all sharing the same starting and terminating sites,
which transforms as $\Gamma$ under all symmetry operations.  The easiest
way to do this is to start with a single path ${\cal P}_\alpha$, such as
a staple, and apply the $\Gamma$ projection operator which is a weighted
sum over all symmetry operations; this yields a single gluon operator in
the $\Gamma$ channel.  Different gluon operators correspond to different
starting paths ${\cal P}_\alpha$.  Using several (typically 3 to 22)
different such operators then produces a matrix of Wilson loop correlators 
$W_\Gamma^{ij}(r,t)$.

Monte Carlo estimates of the $W_\Gamma^{ij}(r,t)$ matrices were obtained
in eight simulations performed on a DEC
AlphaStation 500/333 using an improved gauge-field
action\cite{peardon}.  The couplings $\beta$, input aspect ratios $\xi$,
and lattice sizes for each simulation are listed in 
Table~\ref{table:simparams}.  Our use of anisotropic lattices in which
the temporal lattice spacing $a_t$ was much smaller than the spatial
spacing $a_s$ was crucial for resolving the glue spectrum, particularly
for large $r$.  The couplings in the action depend not only on the QCD
coupling $\beta$, but also on two others parameters: the mean temporal
link $u_t$ and the mean spatial link $u_s$.  Following Ref.~\cite{peardon},
we set $u_t=1$ and obtain $u_s$ from the spatial plaquette.  We use
$a_s/a_t=\xi$, the input or bare anisotropy, in all of our calculations,
accepting the small radiative corrections to the anisotropy as finite
lattice spacing errors which vanish in the continuum limit.

To hasten the onset of asymptotic behaviour, iteratively-smeared spatial
links\cite{peardon} were used in the generalized Wilson loops.
A single-link procedure was used in which each spatial link variable
$U_j(x)$ on the lattice is mapped into itself plus a sum of its four
neighbouring (spatial) staples multiplied by a weighting factor $\zeta$.
The resulting matrix is then projected back into SU(3).  This mapping
is then applied recursively $n_{\zeta}$ times, forming new smeared links
out of the previously-obtained smeared links.  The $(\zeta,n_{\zeta})$
smearing schemes used in the simulations are given in 
Table~\ref{table:simparams}.  Separate measurements
were taken for each smearing; cross correlations were not determined.
The temporal segments in the Wilson loops were constructed from
thermally-averaged links, whenever possible, to reduce statistical noise.

Results for several values of the lattice spacing
were obtained.  Our coarsest lattice ($a_s\sim 0.29$ fm) was used in runs
A, B, and C in order to probe very large quark-antiquark separations.
A large aspect ratio ($\xi=8$)
was needed in order to adequately resolve the correlation functions.
In run A, only on-axis measurements were made.  Run B was done to
reduce uncertainties for the large $r/a_s\geq 9$ measurements.
To verify the restoration of rotational symmetry and to rule out problems
associated with the roughening transition, off-axis 
$(r,r,r)/\sqrt{3}$ measurements were made in run C.  Agreement of energies
obtained using different quark-antiquark orientations on the lattice also
helped identify the continuum $\Lambda$ value corresponding to each level
(there are only discrete symmetries on the lattice).  Runs D, E, F, and G
provided results for different lattice spacings $a_s\sim 0.27,0.22,0.19$,
and $0.19$ fm, respectively.  Only on-axis measurements were made in these
runs.  Runs F and G correspond to the same spatial lattice spacing $a_s$,
but have very different temporal spacings $a_t$.  This provided us with
a measure of the $a_t^2$ errors in our results (our action has
$O(a_t^2,a_s^4)$ errors).  Such information is important for carrying
out the $a_s\!\rightarrow\! 0$ extrapolations.  Run H provided fine-grained
($a_s\sim 0.12$ fm) measurements to assist the continuum-limit
extrapolations of the $\Sigma_g^+$ and $\Pi_u$ potentials and
was useful for verifying the negligible size of quantum corrections to
the lattice anisotropy.  To check the anisotropy, the potentials were
measured for the quark-antiquark axis taken along the very fine-grained
direction and one of the coarser axes was used as the direction of
evolution for the system.  This run was also important for resolving a
slight discrepancy between our $\Pi_u$ results and those\cite{previous,bali}
obtained using the simple Wilson action at $\beta=6.0$.  

The matrices $W_\Gamma^{ij}(r,t)$ were reduced in the data fitting phase
to single correlators and $2\times 2$ correlation matrices using the
variational method.  The lowest-lying glue energies were then extracted
from these reduced correlators by fitting a single exponential and a sum
of two exponentials, the expected asymptotic forms, in various ranges
$t_{\rm min}$ to $t_{\rm max}$ of the source-sink separation.  The 
two-exponential fits were used to check for consistency with the
single-exponential fits, and in cases of favourable statistics, to extract
the first-excited state energy in a given channel.

Three additional runs on small lattices were done to verify that
finite-volume errors in our results were negligible.  We confirmed the
smallness of the $a_s/a_t$ renormalization for two values of the QCD
coupling by extracting the ground state potential from Wilson
loops in different orientations.  The hadronic scale parameter 
$r_0\approx 0.5$ fm
was used to determine the lattice spacing\cite{peardon}. The additive
ultraviolet-divergent self-energies of the static sources were
removed by expressing all of our results with respect to 
$\Sigma_g^+(r_0)$.  Finite-lattice spacing errors were removed
by extrapolating our simulation results for 
$r_0[E_\Gamma(r)-E_{\Sigma_g^+}(r_0)]$ to the 
continuum limit.  These extrapolations were carried out by fitting all
of our simulation results to an ansatz 
$F_{\rm cont}(r) + a_s^4\ F_{\rm latt}(r)$: a ratio of a polynomial
of degree $p+1$ over a polynomial of degree $p$, where $p=1$ or $2$,
was found to work well for the continuum limit form $F_{\rm cont}(r)$,
and $F_{\rm latt}(r)$ was chosen empirically to be a sum of three terms
$1/\sqrt{r}$, $1/r$, and $1/r^2$.  All fits yielded $\chi^2/{\rm dof}$
near unity.  Continuum $\Lambda$ values were easily identified in all
cases but one: we were unable to distinguish between a $\Pi_u^\prime$
and $\Phi_u$ interpretation for the on-axis $E_u^\prime$ level.

\section{Results}

\begin{figure*}[t]
\begin{center}
\leavevmode
\epsfxsize=2.9in\epsfbox[78 224 542 678]{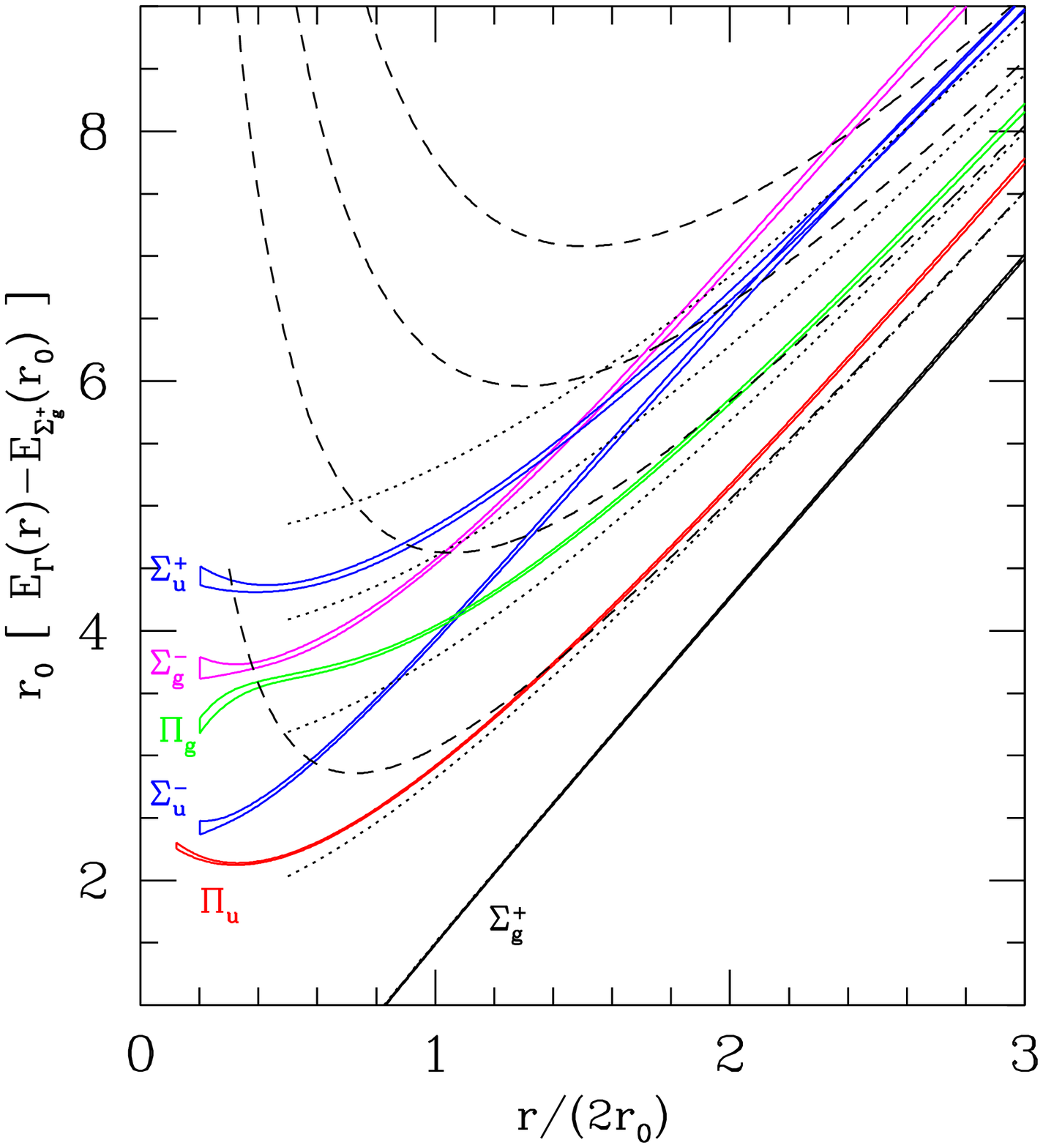}
\epsfxsize=2.9in\epsfbox[78 224 542 678]{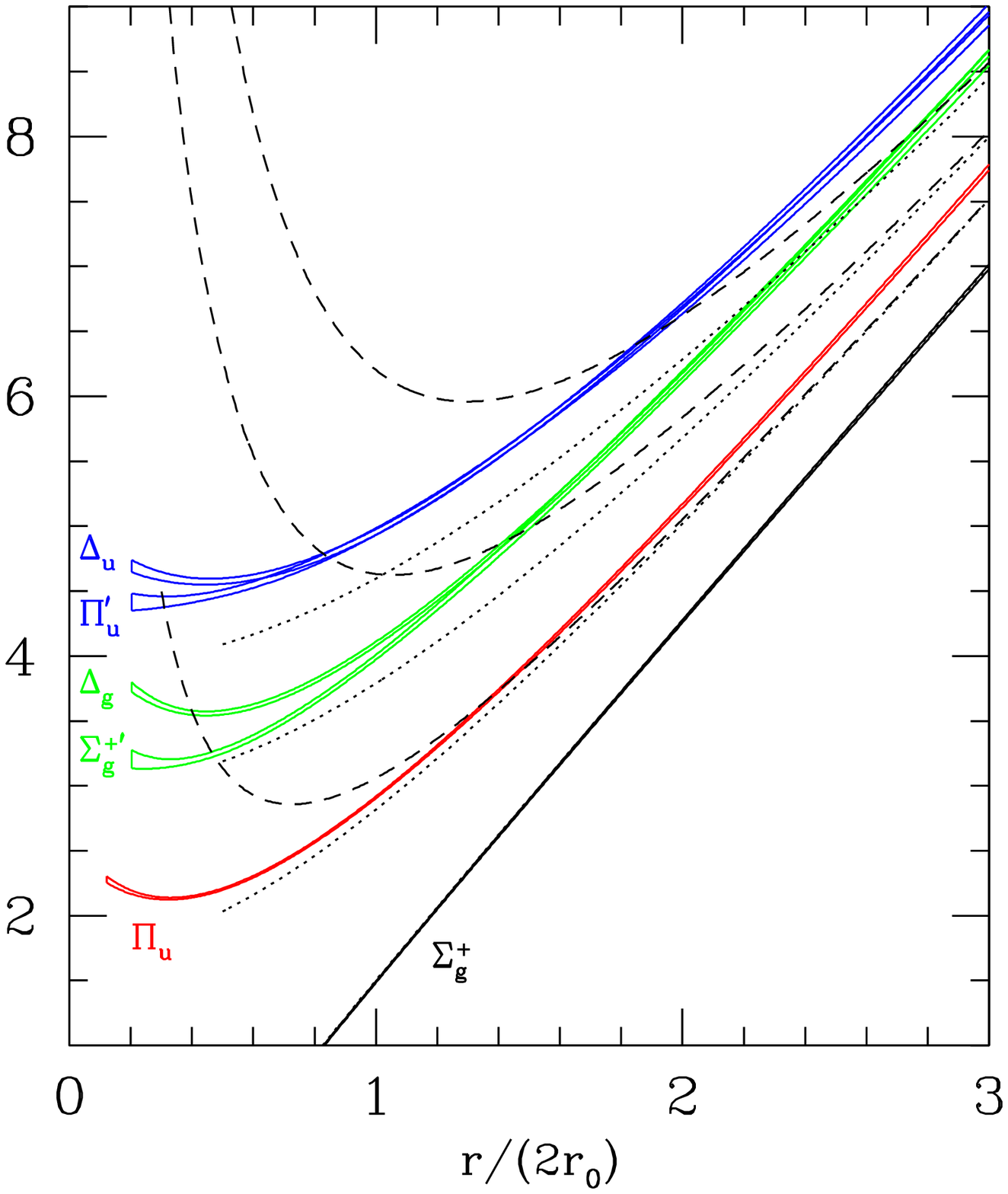}
\end{center}
\caption[figresone]{Continuum-limit extrapolations (with
  uncertainties as indicated) for
  $r_0 [E_\Gamma(r)-E_{\Sigma_g^+}(r_0)]$ against $r/(2r_0)$ for
  various $\Gamma$.  The dashed lines indicate the locations of
  the $m\pi/r$ gaps above the $\Sigma_g^+$ curve for $m=1$, 2, 3, and 4.
  The dotted curves are the naive Nambu-Goto energies in four-dimensions.
  Note that we cannot rule out a $\Phi_u$ interpretation for the
  curve labelled $\Pi_u^\prime$.
  }
\label{fig:res}
\end{figure*}

Our continuum-limit extrapolations are shown in Fig.~\ref{fig:res}.
The ground state $\Sigma_g^+$ is the familiar static-quark potential.
A linearly-rising behaviour dominates the $\Sigma_g^+$ potential once
$r$ exceeds about 0.5 fm and we find no deviations from the linear form
up to 4 fm.  The lowest-lying excitation is the $\Pi_u$.  There is
definite evidence of a band structure at large $r$: the $\Sigma_g^{+\prime}$,
$\Pi_g$, and $\Delta_g$ form the first band above the $\Pi_u$;
the $\Sigma_u^+$, $\Sigma_u^-$, $\Pi_u^\prime/\Phi_u$, and $\Delta_u$
form another band.  The $\Sigma_g^-$ is the highest level at large $r$.
This band structure breaks down as $r$ decreases below 2 fm. In particular,
two levels, the $\Sigma_g^-$ and $\Sigma_u^-$, drop far below their
large-$r$ partners as $r$ becomes small.  Note that for $r$ above 0.5 fm,
all of the excitations shown are stable with respect to glueball decay.
As $r$ decreases below 0.5 fm, the excited levels eventually become
unstable as their gaps above the ground state $\Sigma_g^+$ exceed the
mass of the lightest glueball.

\begin{figure}[t]
\begin{center}
\leavevmode
\epsfxsize=2.8in\epsfbox[78 224 542 678]{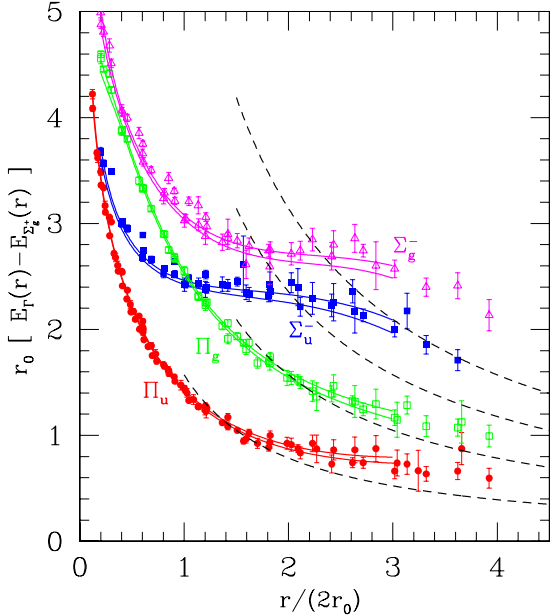}
\end{center}
\caption[figresthree]{Plot of the energy differences
  $r_0 [E_\Gamma(r)-E_{\Sigma_g^+}(r)]$ against $r/(2r_0)$ for
  $\Gamma=\Pi_u$, $\Pi_g$, $\Sigma_u^-$, and $\Sigma_g^-$.  Results
  from all of the simulations are shown; the different symbols correspond
  to the different $\Gamma$ channels.   The pairs of solid curves
  indicate the continuum-limit extrapolations.  The dashed lines show
  the locations of the $m\pi/r$ excitation gaps for $m=1$, 2, 3, and 4.
  }
\label{fig:resC}
\end{figure}

A feature of any low-energy description of a fluctuating
flux tube is the presence of Goldstone excitations associated with the 
spontaneously-broken transverse translational symmetry.  These transverse
modes have energy separations above the ground state given by multiples of
$\pi/r$ (for fixed ends).  The level orderings and approximate degeneracies
of the gluon energies at large $r$ match, without exception, those expected
of the Goldstone modes.  However, the precise $m\pi/r$ gap behaviour is not
observed, as shown in Fig.~\ref{fig:resC}.  The energy differences
$r_0 [E_\Gamma(r)-E_{\Sigma_g^+}(r)]$ for $\Gamma=\Pi_u$, $\Pi_g$, 
$\Sigma_u^-$, and $\Sigma_g^-$ are shown in this figure, along with
their expected Goldstone mode behaviours, indicated by the dashed curves.
For separations less than 2 fm, one sees from Fig.~\ref{fig:res}
that the gluon energies lie well below the Goldstone
energies and the Goldstone degeneracies are no longer observed.  The two
$\Sigma^-$ states are in violent disagreement with expectations from
a fluctuating string.  Note also that our results clearly disagree
with the energies of a Nambu-Goto string naively (ignoring quantization
difficulties) determined in four continuous space-time dimensions.  

These results are rather surprising and cast serious doubts
on the validity of treating glue in terms of a fluctuating string
for quark-antiquark separations less than 2 fm.  Note that such a
conclusion does not contradict the fact that the $\Sigma_g^+(r)$ energy
rises linearly for $r$ as small as 0.5 fm.  A linearly-rising term is not
necessarily indicative of a string: for example, the adiabatic bag model
predicts a linearly-rising ground state much before the onset of
string-like behaviour, even in the spherical approximation\cite{hasenfratz}.
For $r$ greater than 2 fm, there are some tantalizing signatures of
Goldstone mode formation, yet significant disagreements still remain.
To what degree these discrepancies can be explained in terms of
a distortion of the Goldstone mode spectrum arising from the spatial
fixation of the quark and antiquark sources (clamping effect) 
is currently under investigation.  This clamping effect can be confirmed
by comparing our results with the excitation spectrum of a closed flux in
a periodic box in which this effect is absent.  We also plan to study
the $SU(2)$ gluon excitation spectrum in order to ascertain how much of
the spectrum is independent of the gauge group.  Note that it is very unlikely
that the transverse extents of our lattices are causing the discrepancies
with the Goldstone spectrum\cite{michaelB}.  Results for all $r$ up
to 4 fm were insensitive to changes in the transverse size from 0.9
to 3 fm.  As an added check, we also calculated the wave functions for
the first five Fourier modes of the naive Nambu-Goto string; we found
that for $r\sim 4$ fm, the widths of these wave functions did not
appreciably exceed 1 fm.

\section{Hybrid quarkonium}

Another reason for studying the energies of glue in the presence
of a quark-antiquark pair is the likelihood that these energies
will provide insight into the nature of hybrid mesons.  The study of
hybrid mesons comprised of heavy quarks is the natural starting point
in the quest for such an understanding.  A great advantage in studying
heavy hybrid quarkonium is that such systems can be studied not only
by direct numerical simulation, but also using the Born-Oppenheimer expansion.
In this approach, the hybrid meson is treated analogous to a diatomic
molecule:  the slow heavy quarks correspond to the nuclei and the fast
gluon field corresponds to the electrons\cite{hasenfratz}.  The first
step in the Born-Oppenheimer treatment is to determine the energy levels
of the glue (and light quark-antiquark pairs) as a function of the
heavy quark-antiquark separation, treating the heavy quark and antiquark
simply as spatially-fixed color sources.  Each such energy level defines
an adiabatic potential.  The quark motion is then restored by solving
the nonrelativistic Schr\"odinger equation using these potentials.
Conventional quarkonia arise from the lowest-lying potential; hybrid
quarkonium states emerge from the excited potentials.

\begin{figure}[t]
\begin{center}
\leavevmode
\epsfxsize=2.8in\epsfbox[38 224 542 678]{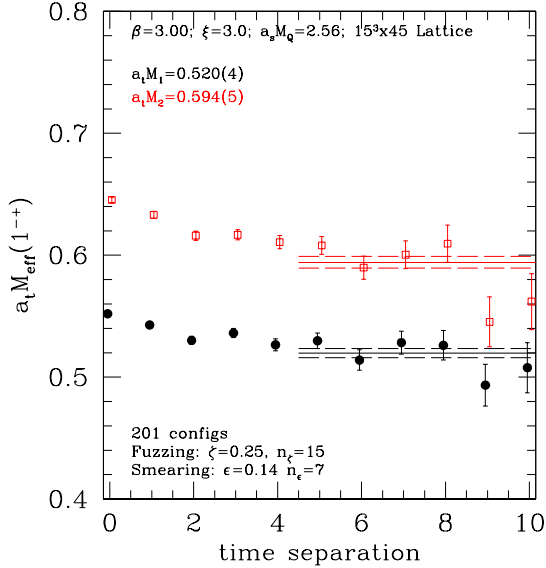}
\end{center}
\caption[figeffmass]{Effective mass plots for the two lowest-lying
  mesons in the exotic $T_1^{-+}$ channel.  Mass estimates from
  fitting the $2\times 2$ correlator matrix elements using their
  expected asymptotic forms are shown.
  The heavy quark propagates according to a spin-independent
  nonrelativistic action.
  }
\label{fig:effmass}
\end{figure}

Once the gluon energies are determined, the Born-Oppenheimer approach
yields the entire leading-order spectrum very easily, in contrast to
direct simulations.  However, the validity of the Born-Oppenheimer
approach relies on the smallness of retardation effects.  One way
of quantifying retardation effects is to compare mass splittings
as determined from the leading Born-Oppenheimer approximation with
those determined from simulations.  A Monte Carlo study with $\beta=3.0$,
$\xi=3$ using a leading-order (but lattice-spacing corrected)
nonrelativistic action for the heavy quark was performed.  The
effective masses for the two lowest-lying mesons in the exotic
$1^{-+}$ channel are shown in Fig.~\ref{fig:effmass}.  The comparison
of mass splittings (for several quark-spin--restored states relative to
the $\Upsilon$) are shown in Table~\ref{table:retard}.  All results
are expressed in terms of the inverse hadronic scale $r_0^{-1}$.  In the
NRQCD simulations, the bare quark mass was taken to be $a_sM_b=2.56$.
The so-called kinetic mass of the $\Upsilon$ was then determined from
its low-momentum dispersion relation.  Half of this mass was used
for the quark mass in the leading Born-Oppenheimer calculation. This
ensured that the $\Upsilon$ kinetic masses were identical in both
calculations.  Assuming that the simulation results do not suffer
significantly from lattice artifacts, one sees that retardation affects
the spin-averaged mass splittings by less than 10\%, validating the
Born-Oppenheimer expansion.

\begin{table}[t]
\setlength{\tabcolsep}{1mm}
\caption[tabone]{Comparison of meson mass splittings as determined from
  the leading Born-Oppenheimer approximation (LBO) and nonrelativistic
  simulations (NRQCD).  The splittings are all taken relative to the
  mass of the $\Upsilon$ and are in terms of $r_0^{-1}$.  The orbital
  angular momentum $L$ and glue energy $\Gamma$ assignments in the LBO
  for each level are indicated, along with the radial quantum number $n$.
  The size of the differences are listed as percentages
  of the simulation results.  
  \label{table:retard}}
\begin{center}
\begin{tabular}{ccclr@{\hspace{2.5ex}}} \hline
  & $nL_\Gamma$ & NRQCD & \hspace{1ex} LBO 
  & \multicolumn{1}{r}{Difference}  \\ \hline\hline
$\chi_b$          & $1P_{\Sigma_g^+}$ & 0.96(1) & 0.872(5)  &  9(1)\% \\
$\Upsilon^\prime$ & $2S_{\Sigma_g^+}$ & 1.30(1) & 1.224(3)  &  6(1)\% \\
$1^{-+}$          & $1P_{\Pi_u}$      & 3.29(5) & 3.166(3)  &  4(2)\% \\
$1^{\prime-+}$    & $2P_{\Pi_u}$      & 4.20(7) & 3.772(2)  & 10(2)\% \\
$0^{+-}$          & $1P_{\Pi_u}$      & 3.51(8) & 3.166(3)  & 10(2)\% \\
$0^{\ast++}$      & $1S_{\Sigma_u^-}$ & 3.56(8) & 3.807(12) &  7(2)\% \\ \hline
\end{tabular}
\end{center}
\end{table}

\section{Conclusion}

The spectrum of gluon excitations in the presence of a static
quark-antiquark pair was comprehensively surveyed for separations
$r$ ranging from 0.1 to 4 fm.  Our results raised serious doubts
on the validity of treating glue in terms of a fluctuating string
for $r$ less than 2 fm.  For $r$ between 2 and 4 fm, some tantalizing
signatures of Goldstone mode formation were observed, but 
discrepancies still remain.  We are currently studying the role
of the spatial fixation or clamping of the quark and antiquark sources
in distorting the Goldstone mode spectrum.
Future studies of the excitation spectrum of the periodically-closed
flux are planned.  Lastly, retardation effects in quarkonium were found
to be sufficiently small to validate the Born-Oppenheimer expansion.
This work was supported by the U.S.~DOE, Grant No.\ DE-FG03-97ER40546.

\end{document}